# A Comprehensive Framework based on Dynamic and Steady-State Analysis to Evaluate Power System Resiliency to Extreme Weather Conditions

Giritharan Vijay Iswaran
School of Electrical, Computer, and Energy Engineering
Arizona State University
Tempe, Arizona, USA
gvijayis@asu.edu

Ramin Vakili
School of Electrical, Computer, and Energy Engineering
Arizona State University
Tempe, Arizona, USA
rvakili@asu.edu

Mojdeh Khorsand
School of Electrical, Computer, and Energy Engineering
Arizona State University
Tempe, Arizona, USA
mojdeh.khorsand@asu.edu

*Abstract*—Power system robustness against high impact low probability events is becoming a major concern. To depict distinct phases of a system response during these disturbances, an irregular polygon model is derived from the conventional trapezoid model and the model is analytically investigated for transmission system performance, based on which resiliency metrics are developed for the same. Furthermore, the system resiliency to windstorm is evaluated on the IEEE reliability test system (RTS) by performing steady-state and dynamic security assessment incorporating protection modelling and corrective action schemes using the Power System Simulator for Engineering (PSS®E) software. Based on the results of steady-state and dynamic analysis, modified resiliency metrics are quantified. Finally, this paper quantifies the interdependency of operational and infrastructure resiliency as they cannot be considered discrete characteristics of the system.

*Index Terms*-- Dynamic security assessment, extreme weather condition, irregular polygon model, operational and infrastructure resiliency.

## I. INTRODUCTION

An electric power system is a highly complex, nonlinear, and non-convex system whose uninterrupted operation is of paramount importance for any modern society. Severe events can create disturbances in power systems that lead to operation interruptions and even major cascading outages and blackouts if proper preventive actions are not being considered to make the system robust against such events. The Texas outage due to a snowstorm which created a major impact on people's life [1]-[2] is an example of such outages. Such events are referred to as high impact low probability events which include natural disasters and man-made disasters.  In recent years, power system resiliency has been a nascent subject attracting many scholars [3]-[5] to understand its operational and infrastructural characteristics.

Based on a study conducted in [6], among different scenarios, extreme weather events have been constantly reported causing huge impacts on system operation. With the increasing trend of power system outages that are induced by extreme weather conditions, enhancing power system resiliency to such events is gaining more attention in the research community. Apart from the reliability metrics, numerous resiliency metrics have been developed recently to quantify the network robustness against windstorms as detailed in [7]. Diverse quantifications are considered in the literature such as analyzing system adaption and disruption [8]-[10], and network recovery both at operational and structural levels [11]. Resiliency incorporates long and short-term planning and adjustments during event progression and post-shock periods. Defensive islanding, effective remedial action design, and system hardening are among long-term planning measures, whereas short-term measures include event forecasting, accurate assessment of the availability of assets, as well as the implementation of effective preventive and corrective actions.

Resiliency evaluation approaches, such as triangle and trapezoid methods, are used extensively to represent infrastructural (e.g., transmission corridor status) and operational (e.g., load and generator shedding) performances in [12] and [15]. The resiliency trapezoid model incorporates ramping decrement in performance indictor during an event and modifies the resiliency triangle assertion of impulsive drop during the shock absorption. Moreover, prior work considered a linear representation of resilience indicators as the event propagates and the system recovers. All these prior analyses focus only on the steady-state response of the system. However, power system security during extreme events is highly dependent on its dynamics and protection systems. Prior work fully ignores power system dynamics and protection system behaviour while analyzing power system resiliency to extreme events. In order to bridge the gap between actual system performance and estimated system performance, i.e., using steady-state analysis, this paper focuses on evaluating system responses based on steady-state and dynamic analysis with protective relay modelling. The overall result obtained is coined as an aggregated result. The term 'aggregated result' is used to address the combined results of steady-state and dynamic analysis in the upcoming sections.

Based on the fragility model for the transmission assets extracted from [7], the impact of windstorms on system performance, both in the steady-state and transient period, is analyzed and new resilience metrics are proposed in this paper.

The contributions of this paper include:

(1) Development of a new framework to evaluate power system resiliency by assessing the transient stability of the system after critical changes in the system assets during extreme weather events along with the evaluation of the steady-state condition of the system. Moreover, critical protective relays are included in transient stability analysis. This contribution is of critical importance as the power system response during extreme events is usually governed by the dynamic response of generators, control systems, and protection system behaviour.

(2) Highlighting the importance of interdependence analysis of the infrastructural and operational resiliency.

The rest of the paper is organized as follows. Section II briefly discusses the existing resiliency evaluation methods and provides the modified approach for system resiliency analysis considering steady-state and dynamic conditions. Section III provides the mathematical formulations for quantifying power system resiliency based on the steady-state and aggregated result (steady-state and dynamic analysis). Section IV introduces the test system and discusses the numerical results. Conclusions are provided in Section V.

## II. QUANTITATIVE EVALUATION OF POWER SYSTEM RESILIENCY

Power system resiliency is usually categorized as infrastructure and operational resiliency. The infrastructure performance indicators represent power system component status during events. The operational resiliency indicators quantify various factors such as the percentage of load served, generation availability, and frequency deviation. The first part of this section tends to briefly discuss the infrastructure and operational resiliency based on linear models such as the trapezoid method discussed in [7], whereas the latter section proposes an irregular polygon model for measuring system resiliency. This model is referred to as the aggregated model in this paper.

### A. Infrastructure and Operational Resiliency: Trapezoid Model

The crucial assets of the electric power systems involve generators, transmission lines, transformers, and distribution lines. Responses of these devices during the event are monitored and are collectively referred to as the power system response. During any event, the power system response framework indicating the operational and infrastructural performance with respect to time can be schematized and depicted as shown in Figure 1. In general, the system response can be categorized into four phases, namely, intact, absorption and adaption, recuperation, and recovered operation (RO). The first phase, known as the intact phase, is a pre-contingency state, where the system operation is undamaged or unaffected. The performance indicator at this stage is ideally 100% and the system remains in an undamaged state until $t = t_0$. In the absorption and adaption phase, the system resiliency indictor decreases during event progression until $t_{ee}$ and the system operates in a new state immediately after the event. The operational and infrastructure indicators remain at the post-disturbance state until the time of $t_{sr}$.

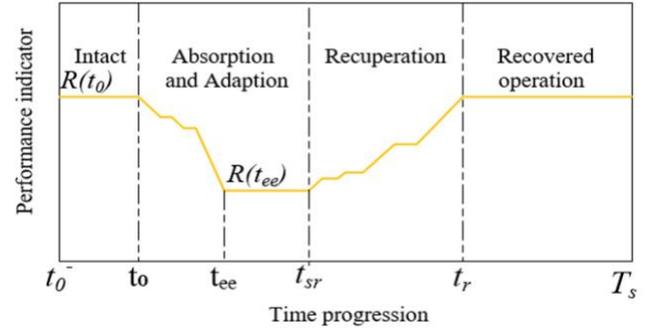

Figure 1. Modified multiphase trapezoid curve

In the third phase, restoration begins at $t_{sr}$, and the system tries to reach its pre-contingency state. Both the infrastructure and operational indicators slowly recover until $t_r$. The final phase or the recovered operation indicates the measure of infrastructure and operational indicators during the post-recovery state. Unlike infrastructure resiliency, operational resiliency recovery is influenced by demand, generation availability, asset operational capability, and implementation of effective corrective actions. System resiliency during the post-recovery period can indicate an improved/deteriorated performance or the same pre-contingency performance of the system.

### B. System Resiliency: Irregular Polygon Model

The resiliency measure depends on asset outages and human intervention during the recovery phase. A linearized irregular polygon model [15] is defined in this section for defining the system resiliency indicator as shown in Figure 2.

Similar to the infrastructure/operational resiliency assessment (IORA), the system level performance has four phases, namely, intact, absorption, recuperation, and recovered operation. However, unlike the recuperation phase in IORA, the recuperation phase in system-level performance has three states, namely, operational recovery, quasi recovered state, and infrastructure and operational (I&O) recovery. The infrastructure recuperation phase is a highly laborious and time-consuming phase defining the recovery rate of the system after the disruption period. However, during this asset repairing period, the system operators can regain a part of the lost load and disconnected generators by implementing corrective operational actions, e.g., generator re-dispatch, power import from (export to) another area; this stage is defined as operational recovery in this paper. Starting from $t_{o\_sr}$, this state prolongs until $t_{q\_sr}$. During the next state, known as the quasi-recovered state, a part of disconnected customers/generators and islanded buses/areas are regained back and the system reaches a new operational state with a lower performance indicator compared to the targeted value. This state lasts until $t_{io\_sr}^{(s)}$. I&O recovery state, being the third state, interprets the system performance improvement by recovering the infrastructural and operational characteristics of the network until $t_r^{(s)}$. Note that there can be multiple states of operational recovery as more assets are repaired and new operational actions are implemented (Figure 1).

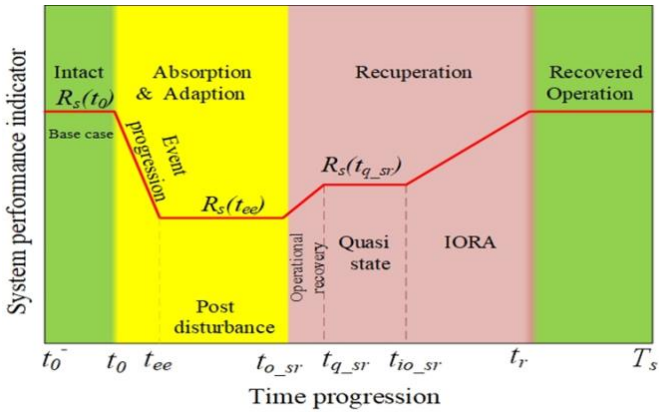

Figure 2. Irregular polygon model for system resiliency

## III. RESILIENCY FORMULATION AND SYSTEM ANALYSIS

Based on the IORA, infrastructural and operational performance indicators such as disruption rate, preparation time, recovery rate, and the area under the curve are briefly defined in this section.

### A. System Resiliency Metrics

The numerical metrics used to study the system response during extreme conditions can be expressed as shown in Table I using the notations of Figure 1.

TABLE I. STEADY STATE RESILIENCY METRICS

| Metrics | Formulation | Unit |
|---|---|---|
| Intact period | $t_0 - t_0^-$ | Hours |
| Disruption rate $D_r$ | $\dfrac{R(t_0) - R(t_{ee})}{t_{ee} - t_0}$ | Units/Hours |
| Preparation time | $t_{sr} - t_{ee}$ | Hours |
| Recovery rate $\Re$ | $\dfrac{R(T_s) - R(t_{ee})}{t_r - t_{sr}}$ | Units/Hours |
| Recovery time $\tau_r$ | $t_r - t_{sr}$ | Hours |
| $\Lambda_t$ | Area under curve | |
| Absorption time | $t_{sr} - t_0$ | Hours |

### B. Fragility Modelling of Transmission Network

Fragility analysis is a probabilistic approach to determine the reliability of transmission networks (including transmission lines and towers) during extreme weather conditions and can be generally represented as (1).

$$\vartheta(v_i) = \begin{cases} 0, & if\ v < v_{cr} \\ \vartheta(v), & if\ v_{cr} \leq v \leq v_{br} \\ 1, & if\ v > v_{br} \end{cases} \quad (1)$$

Where $\vartheta(v)$ is the failure probability of the transmission network for any wind speed of $v$ m/s and $\vartheta(v_i)$ is the failure probability of the transmission network for the wind speed of $v_i$ m/s when the wind speed is bounded within the critical wind speed ($v_{cr}$) and the breakdown wind speed ($v_{br}$). For a wind speed below $v_{cr}$, the infrastructure and operation of the transmission network remain unaffected (i.e., zero failure probability). Whereas, for a wind speed above $v_{cr}$ the reliability of the transmission network is altered based on the failure probability function. Wind speed above $v_{br}$ collapses the structural and operational template of the network (i.e., 100% failure probability). Based on the fragility data, the status of the transmission corridor can be decided using a uniformly distributed random number ($R \in (0,1)$. The transmission network goes offline if the failure probability is greater than the generated random number (i.e., $\vartheta(v) > R$ ). The transmission network recovery depends on the mean time to repair and the severity of the weather intensity.

The approach that has been followed in this paper to define and evaluate the resiliency metrics is depicted in Figure 3. The process involves conducting steady-state and dynamic analysis of the system at each hour of event propagation to calculate the proposed metrics. Transient stability analysis with modelling protection schemes—including distance relays, underfrequency load shedding, and undervoltage load shedding schemes—as well corrective actions are carried out at each hour of the event period. The transient stability studies evaluate system dynamic stability and identify the lines that may be tripped (due to overload conditions or unstable power swings) and the loads/generators that are shed during the transient period. The results of transient stability (including line tripping and generator/load shedding) are utilized to make the related updates in the steady-state system topology at every time step. A steady-state analysis of the updated system involving AC optimal power flow (AC-OPF) is performed with minimization of the bus load adjustment (amount of shed load) as the objective function.

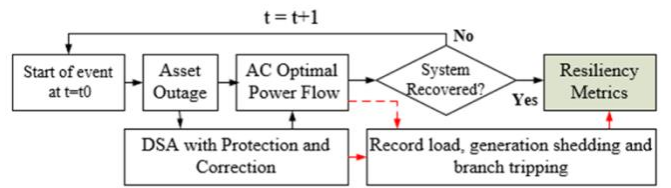

Figure 3. The operational approach in determining resiliency metrics

## IV. TEST SYSTEM AND RESULTS

The IEEE reliability test system is used for testing the proposed framework using the software tool PSS®E version 35.2. In this section, the reliability test case system, system dynamic and fragility model are briefly discussed, and the steady-state and aggregated results are analyzed and compared as well. Besides, the importance of the aggregated result is brought to notice in the system resiliency study considering stability as one of the key factors of discussion.

### A. IEEE Reliability Test Case 1996

The IEEE reliability 1996 test (RTS'96) system proposed in [14] is considered for analyzing and quantifying resiliency metrics in this study. RTS is a three-area 73 bus system having a total of 99 conventional generators, 104 transmission lines, 16 two-winding branches, and 51 load buses. The total generation capacity of the system is 14,550 MW and the peak load demand is 8725.26 MVA. The mean time to repair (MTTR) for transmission lines is considered one-fifth of that of transmission towers. Following are the assumptions considered throughout the period of the simulation study.

1) The resiliency study is carried out for a power system against a windstorm affecting only area 1. The event initiates at t = 51 and prolongs for a period of 25 hours.

2) The simulation is carried out for 400 hours assuming the availability of the labour force after the windstorm event. The simulation hours can extend up to a period of 600 to 750 hours based on the storm intensity level.

3) The study is carried out only for transmission level and the bus loads are modelled as lumped constant power load model. The corresponding hourly bus loads are scheduled based on the active and reactive power demand during the winter season on weekdays/weekends.

4) The fragility model of the transmission networks is designed assuming that the 138 kV and 230 kV systems as normal (less robust) and more robust corridors, respectively.

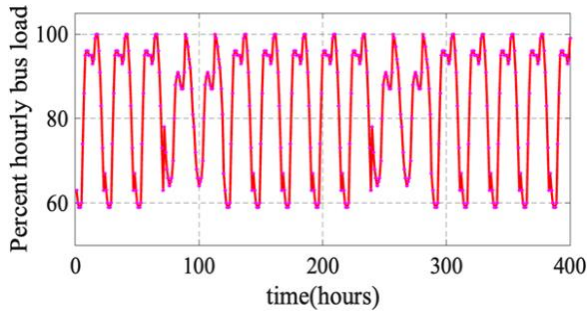

Figure 4. Winter load profile

5) Any outages that happened in the system during the event progression are not recovered back during the next simulation hour within the event period for steady-state analysis. However, the assets lost during transient stability analysis are regained back within the next 20 minutes despite this recovery might take an even longer period in real-time scenarios. During event progression, the long-term thermal rating (24-hour rating) and short-term thermal rating (15-min rating) for the branches are considered for steady-state and dynamic analyses, respectively.

The hourly load profile and wind speed variation during the event progression are depicted in Figures 4 and 5, respectively. The fragility model representing transmission line/tower failure probability with respect to different wind speeds for less robust and more robust systems is extracted from [7] and is shown in Figure 6.

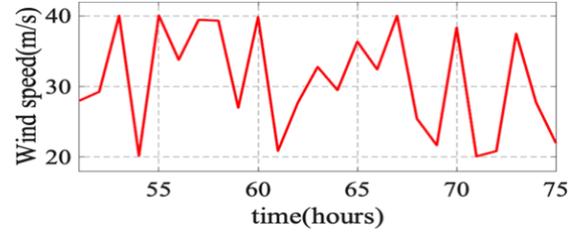

Figure 5. Wind speed profile

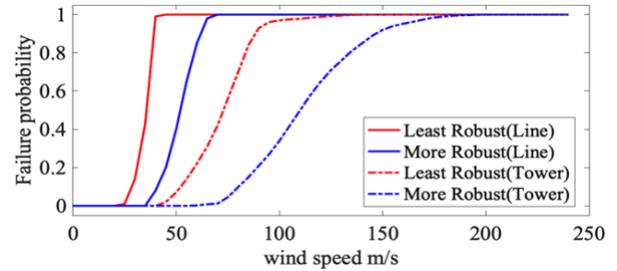

Figure 6. Fragility curve of transmission corridor extracted from [7]

### B. Dynamic Modeling of System Components

The GENROU model from the PSS®E software model library is used to model the synchronous generators in the transient stability studies.

TABLE II. LOAD SHEDDING AND DISTANCE RELAY PARAMETERS

| UNDER-FREQUENCY LOAD RELAY | | | UNDER-VOLTAGE LOAD RELAY | | |
|---|---|---|---|---|---|
| Frequency (Hz) | Load Shed (%) | Time delay (sec) | Voltage (p.u.) | Load Shed (%) | Time delay (sec) |
| 60.0 | - | - | 1.00 | - | - |
| 59.5 | 10 | 0.05 | 0.90 | 30 | 0.75 |
| 59.2 | 20 | 0.05 | 0.85 | 50 | 1.00 |
| 59.8 | 20 | 0.05 | - | - | - |
| DISTANCE RELAY | | | | | |
| Zones | | (%) distance | | Time delay (cycles) | |
| 1 | | 90 | | 0 | |
| 2 | | 120 | | 15 | |
| 3 | | 220 | | 90 | |

GENROU represents a round rotor generator model and its corresponding exciter model used is EXST1 (1992 IEEE ST1 model). The constant power bus loads are modelled using the PSS®E model IEEEBL1. HYGOV, GGOV1, and TGOV models have been employed for the machine governor models. Undervoltage/frequency load shedding (UVFLS) relays are developed using PSS®E user-defined model UVUFBLU1 and are included in the studies. DISTR1 model from the PSS®E software model library is used to model distance relays (mho relay). Parameters for load shedding relay and distance relay models are selected based on the NERC rules and regulations [13] and are tabulated in Table II respectively. The breaker delays for UVLS and UFLS models are 0.05 and 0.02 seconds, respectively.

*C. Steady-State Results*

AC optimal power flow module of PSS®E is used for steady-state analysis with the aim of minimizing the load shedding at each bus. The steady-state infrastructural and operational resiliency indicators based on the availability of generators, served bus loads, and connected transmission lines are plotted in Figures 7, 8, and 9, respectively.

There are 99 generation units in the test case, and during the event progression, a total of 15 units become offline because of bus islanding due to line outages, decreasing the available generation capacity from 14,550 MW to 13530 MW at the end of the event. As a result of the isolation of load buses due to the windstorm effect, approximately 4% of the load is shed.

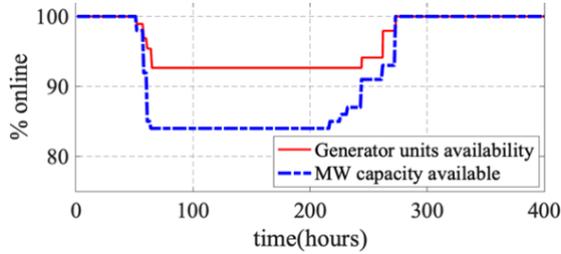

Figure 7. Hourly generator units and active power capacity availability

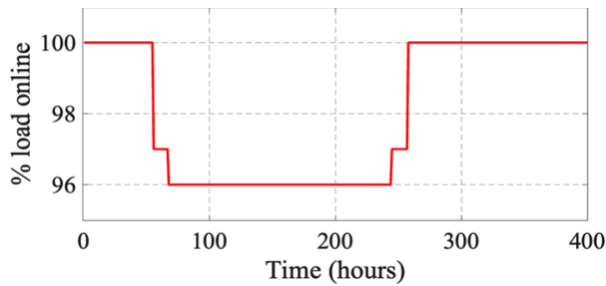

Figure 8. Percentage of load connected

In Figure 9, there are two curves representing the operational statuses of the transmission lines (solid line) and the infrastructure availability of the transmission lines (dashed line). The dashed line represents the number of transmission lines that were not damaged by the windstorm whereas the solid line represents the operational status of all transmission lines. A transmission branch becomes operationally unavailable due to bus islanding or complete damage because of the event. However, the line is infrastructurally available if not damaged by the event. The recovery time for the towers and lines that are damaged depends on their mean time to repair rates obtained from the IEEE RTS data.

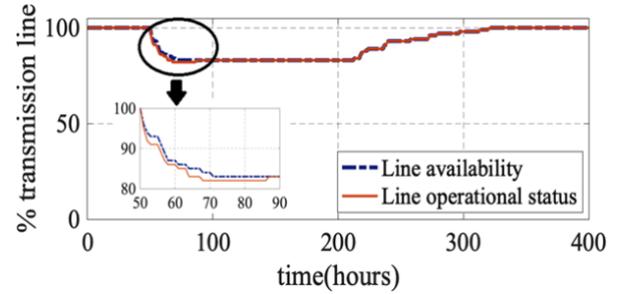

Figure 9. Percentage of transmission line connection/availability status

The curves in Figure 9 match the irregular polygon model discussed in section II.A. The resiliency metrics provided in Section III. A for the generation units, loads, and transmission corridors are calculated and are provided in Table III. In Table III, IR and OR refer to infrastructure resiliency and operational resiliency, respectively.

TABLE III. STEADY-STATE RESILIENCY METRICS

| Metrics | Generator | | Branch | | Load |
|---|---|---|---|---|---|
| | IR | OR | IR | OR | OR |
| $D_r$ (%/hr) | -0.667 | -0.306 | -0.75 | -0.71 | -0.167 |
| $\Re_r$ (%/hr) | 0.301 | 0.023 | 0.077 | 0.148 | 0.307 |
| $\tau_r$ (hr) | 47 | 47 | 234 | 108 | 13 |
| $\Lambda_t$ (p.u) | 0.952 | 0.994 | 0.918 | 0.92 | 0.981 |

*D. Aggregate Results*

Aggregated results include the steady-state and dynamic analysis results. Transient stability analysis based on rotor angle stability is carried out for a time period of 15 seconds. The branch rating is set to STE (15-min emergency rating). The branch status is updated (disturbance is introduced) 1 second after the study initiation time. The step size for the simulation is 0.0083 s (which is half of the 60 Hz cycle period). Dynamic performance of the system in terms of transient load shedding and generator shedding around the corresponding simulation hour is plotted in Figures 10, and 11, respectively.

Both the generator and load curves have dips at certain hours during the simulation. These dips in the curves are due to the transient generator and load shedding which can be recovered after a period of 20 minutes based on the above assumption. Figures 10 and 11 represent the aggregated results and reveal the actual number of generators and loads that become offline during the event. It is observed that a load drop of 83.6 % occurs at simulation hour of 237 because of reconnecting generator bus 122. These dips in the generator and load curves (generator/load shedding) are not observed in the steady-state analysis. Thus, the steady-state analysis alone is not sufficient for having a realistic assessment of the system resiliency since the system stability and the dynamic response of the system are important to monitor. The following paragraphs brief the importance of dynamic analysis after a major disturbance by providing an instance where an intense outage happened causing system instability during the study.

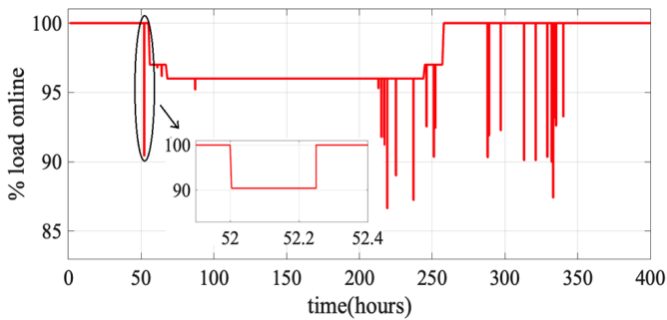

Figure 10. Percentage load online-aggregated model

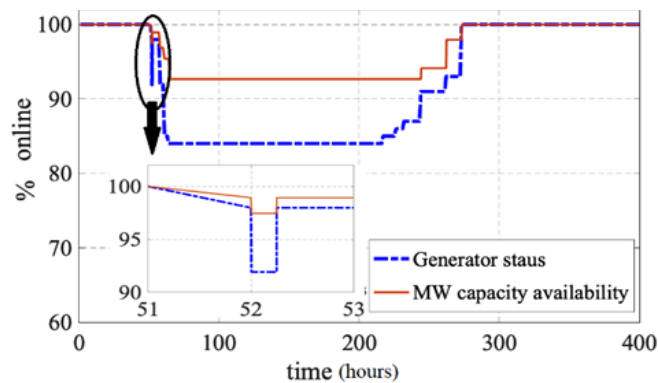

Figure 11. Percentage generators and active power capacity availability-aggregated model

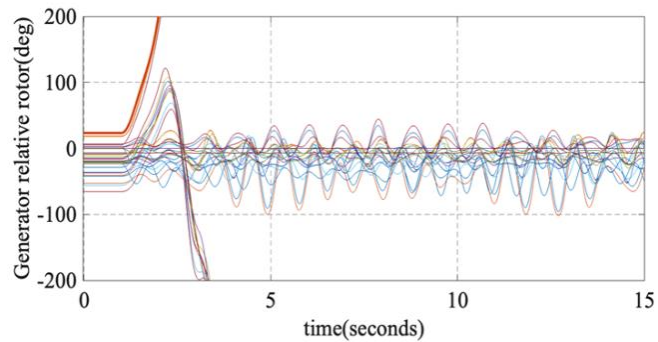

Figure 12. Relative rotor angle of machines at hour 52 without corrective action

During transient stability studies, the stability of the system is evaluated by monitoring the relative rotor angles, voltages and bus frequency of the connected machines. One such instance during the simulated event that can jeopardize system stability occurred at hour 52. If proper corrective actions were not taken within the effective time, the tripping of the line between Arnold 230 kV and Aston 230 kV buses at hour 52 can jeopardize system stability. Due to the storm effect at hour 52, the transmission line between buses 114 and 117 is damaged. Due to this tripping action, the machines at bus 122 lose synchronism and this action of losing synchronism is observed at generators connected to buses 114, 115, 116, and 123 in area I. Because of unstable power swings at generator buses 325 and 223, the two tie branches 325-121 and 318-223 are tripped by distance relay operation causing line overloading. It is also further observed that almost all machines located in area II and area III lost their synchronism at this hour.

The relative rotor angles of the generators at hour 52 are shown in Figure 12. Actions such as generators tripping at bus 122 by out of step relay can prevent the system instability in this case. This action makes the system stable and prevented distance relay misoperation and the consequent tripping of the tie lines. The relative rotor angles of the generators after applying the corrective actions are shown in Figure 13. As can be seen in this figure, no generator loses its synchronism with respect to the rest of the system.

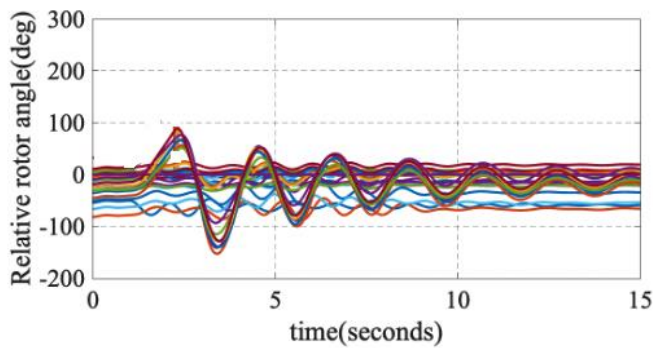

Figure 13. Relative rotor angle of machines at hour 52 with corrective action

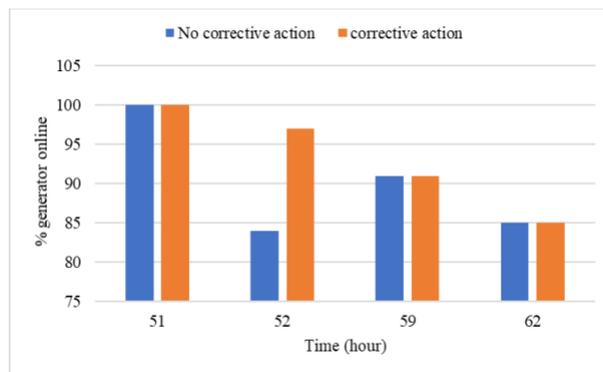

Figure 14. Percentage of the generators online in the two cases of with and without implementing corrective actions

The generator sheddings with and without corrective actions at hour 52 are compared in Figure 14. As can be seen, with corrective actions being implemented at hour 52, the percentage of the generators remaining online increased substantially. Similarly, the load shedding can be compared for the mentioned cases. At hour 52, the load shedding has reduced from 94.5% to 8.1% by implementing the corrective actions, which shows that the remedial actions prevent the system from collapsing and losing all its load. The system resiliency metrics based on the aggregated results of the system are computed and tabulated in Table IV. In Table IV, IR and OR refer to infrastructure resiliency and operational resiliency, respectively.

TABLE IV. Aggregated Resiliency Metrics

| Metrics | Generator | | Branch | | Load |
|---|---|---|---|---|---|
| | IR | OR | IR | OR | OR |
| $D_r$ (%/hr) | -0.667 | -0.306 | -0.75 | -0.71 | -0.16 |
| $\Re_r$ (%/hr) | 0.301 | 0.023 | 0.077 | 0.148 | 0.021 |
| $\tau_r$ (hr) | 47 | 47 | 234 | 108 | 187 |
| $\Lambda_t$ (p.u) | 0.918 | 0.96 | 0.918 | 0.92 | 0.976 |

Comparing the disruption, recovery rate, and area under the curve metrics of the aggregate results, it is observed that their values are less than that in the steady-state studies since steady-state studies do not consider system instability, transient load/generator shedding, and line tripping. The existing steady state (trapezoid) resiliency model is not sufficient for analysing system resiliency as it neglects dynamic stability, transient load shedding and transmission line tripping. These issues are addressed in dynamic studies and specific corrective actions are implemented to make the system operate in stable condition. Besides, the multiple stages of recovery can be visualized in the irregular polygon model, which cannot be shown in the current trapezoid model.

## V. Conclusion

This paper analyzes the steady-state and dynamic effects of windstorms on power systems. The studies are conducted on the IEEE reliability test system 1996. An irregular polygon model has been developed based on the extension of the conventional multiphase trapezoid model. The system-level resiliency model is introduced and is quantified based on the resiliency metrics including disruption rate, recovery rate, recovery time, absorption time, and area under curve metrics. The system-level recovery rate formulation has been modified based on the existing trapezoid model considering the effect of quasi recovery state on the model. Disruption rate, recovery rate, and area under curve metrics are modified and proposed based on the aggregate results obtained from the steady-state and transient stability results. Moreover, the importance of including transient stability analysis in power system resiliency studies and its effects on evaluating the system's operational status such as load shedding and generation shedding is evaluated in this paper. The rotor angle stability of the system is evaluated throughout the event period, and the impact of implementing proper corrective actions on preventing rotor angle instability and improving the resiliency metrics of power systems is analyzed. Finally, the resiliency metrics for steady-state and aggregate results are computed and compared. It can be observed that the area under the curve, disruption rate, and recovery rate metrics are highly impacted in the dynamic analysis due to the presence of dips in load and generator status curves. In addition to the existing steady-state analysis, the dynamic analysis tool paved the way for realizing the system operational status that can closely mimic the actual system performance. Future work progresses on evaluating the impact of the high penetration level of renewable resources on the resiliency of the power grids.